\documentclass[12pt,preprint]{aastex}

\newcommand{\degree}{$^{\circ}$}

\newcommand{\HII}{H\,{\sc ii}}
\newcommand{\kms}{km~s$^{-1}$}

\slugcomment{Accepted for publication in ApJ}

\shorttitle{YSO near IRAS 07029-1215}
\shortauthors{Forbrich et al.}

\begin{document}

\title{An Extremely Young Massive Stellar Object \\ near IRAS 07029-1215}

\author{J. Forbrich\altaffilmark{1}, K. Schreyer, B. Posselt\altaffilmark{2}, R. Klein\altaffilmark{2}}
\affil{Astrophysikalisches Institut und Universit\"ats-Sternwarte, Friedrich-Schiller-Universit\"at, Schillerg\"a\ss 
chen 2-3, D-07745 Jena, Germany}

\and

\author{Th. Henning}
\affil{Max-Planck-Institut f\"ur Astronomie, K\"onigstuhl 17, D-69117 Heidelberg, Germany}

\altaffiltext{1}{Max-Planck-Institut f\"ur Radioastronomie, Auf dem H\"ugel 69, D-53121 Bonn, Germany; forbrich@mpifr-bonn.mpg.de}
\altaffiltext{2}{Max-Planck-Institut f\"ur extraterrestrische Physik, Giessenbachstra{\ss}e, D-85748 Garching, Germany}

\begin{abstract}
In the course of a comprehensive mm/submm survey of massive star-forming
regions, the vicinities of a sample of 47 luminous IRAS sources were closely investigated with SCUBA and IRAM 
bolometers in order to 
search for massive protostellar candidates. A particularly interesting object has been found in the
surroundings of the bright FIR source IRAS 07029-1215. Follow-up line observations show that the object is cold, that it has a massive envelope, and that it is associated 
with an energetic molecular outflow. No infrared point source has been detected at its position. Therefore, it is a very good candidate for a member of the long searched-for group 
of massive protostars.
\end{abstract}

\objectname[IRAS 07029-1215]{}


\section{Introduction}
The study of the earliest stages of massive star formation (MSF) is one of the most exciting subjects of present research in this field. 
Whether the dominant 
formation process for massive stars is disk accretion (see e.g. \citealp{jia96}) or coalescence \citep{bon98} is still an 
open question. The presence of outflows in regions of massive star formation favours, however, the first scenario. Similarly to previous
surveys (e.g. \citealp{beu02b}, \citealp{mue02} or \citealp{hun00}), we have searched for MSF candidates located near bright IRAS sources using SCUBA and IRAM bolometers (\citealp{pos03} and Klein et al. 2003, {\sl in prep.}). Our focus was on objects unrelated to the IRAS sources themselves, looking for the earliest evolutionary stages not yet seen in the near- to mid-infrared wavelength regime. Even if we do not yet know what the observational features of massive protostars are \citep{eva02}, their invisibility at NIR/MIR wavelengths is a defining criterion.
Near IRAS~07029-1215, an object with an IRAS luminosity of $L(IRAS)=1700$~L$_{\odot}$ \citep{hen92} at its distance of $d=1$~kpc, we discovered a deeply embedded object 
(UYSO~1, for \textsl{Unidentified Young Stellar Object~1}), powering a high-velocity bipolar CO outflow. 
The IRAS source is located in the bright \HII{} region S~297, illuminated by the B1II/III star
HD~53623 \citep{hou88}. UYSO~1 is located very close to the edge of a dark cloud.

\section{Observations}
CO($J=3 \to 2$) line mapping of UYSO~1 was carried out with the James Clerk Maxwell Telescope (JCMT)
\footnote{JCMT is operated by the Joint Astronomy Center on behalf of the Particle Physics and Astronomy Research 
Council of the United 
Kingdom, the Netherlands Organization for Scientific Research, and the National Research Council of Canada.}
on Mauna Kea, Hawaii, in October 1999. The facility receiver B3 (315-373 GHz) was used as frontend, the Digital 
Autocorrelation Spectrometer
(DAS) as backend. With a total bandwidth of 250 MHz, centered on v$_{lsr} = +12$~\kms{}, the maps were obtained in 
position-switch mode 
with an 
off-position 10\arcmin{} to the east. Map sampling was at intervals of 10\arcsec{}, the total on+off integration time was 10 min, the beam efficiency $\eta = 0.63$, and the beam size was 14\arcsec{}. Additionally, SCUBA continuum observations at $\lambda = 450$~$\mu$m and $\lambda = 850$~$\mu$m were performed with a total integration time of 46 minutes. Here, beam sizes were 8\arcsec{} and 14\arcsec{}, respectively. Based on six different field scans with different chop throws, noise was minimised using the Emerson~II technique \citep{eme95}. In March 2003, additional CS($J=2 \to 1$), CS($J=5 \to 4$), SiO($J=2 \to 1$) and especially H$_2$CO($J=3_{03}-2_{02}$)/($J=3_{22}-2_{21}$) observations were obtained at the IRAM 30m telescope using the B100, B230, A100 and
A230 receivers. The raster map sampling was the same as for the JCMT observations, though a smaller range of offset
positions was observed. The center position is shifted southward by 6.4\arcsec{} when compared to the JCMT map. Data was obtained in position switch mode, integration time at each offset was 6~min. All line observations together with the observing parameters are summarized in Tab.~\ref{t0}.

We used the CLASS software developed by the Grenoble Astrophysics Group for the line data reduction, including zero-order baseline subtraction, and GAIA for the SCUBA data.

\section{Results}
\begin{figure}
\rotatebox{-90}{
\plotone{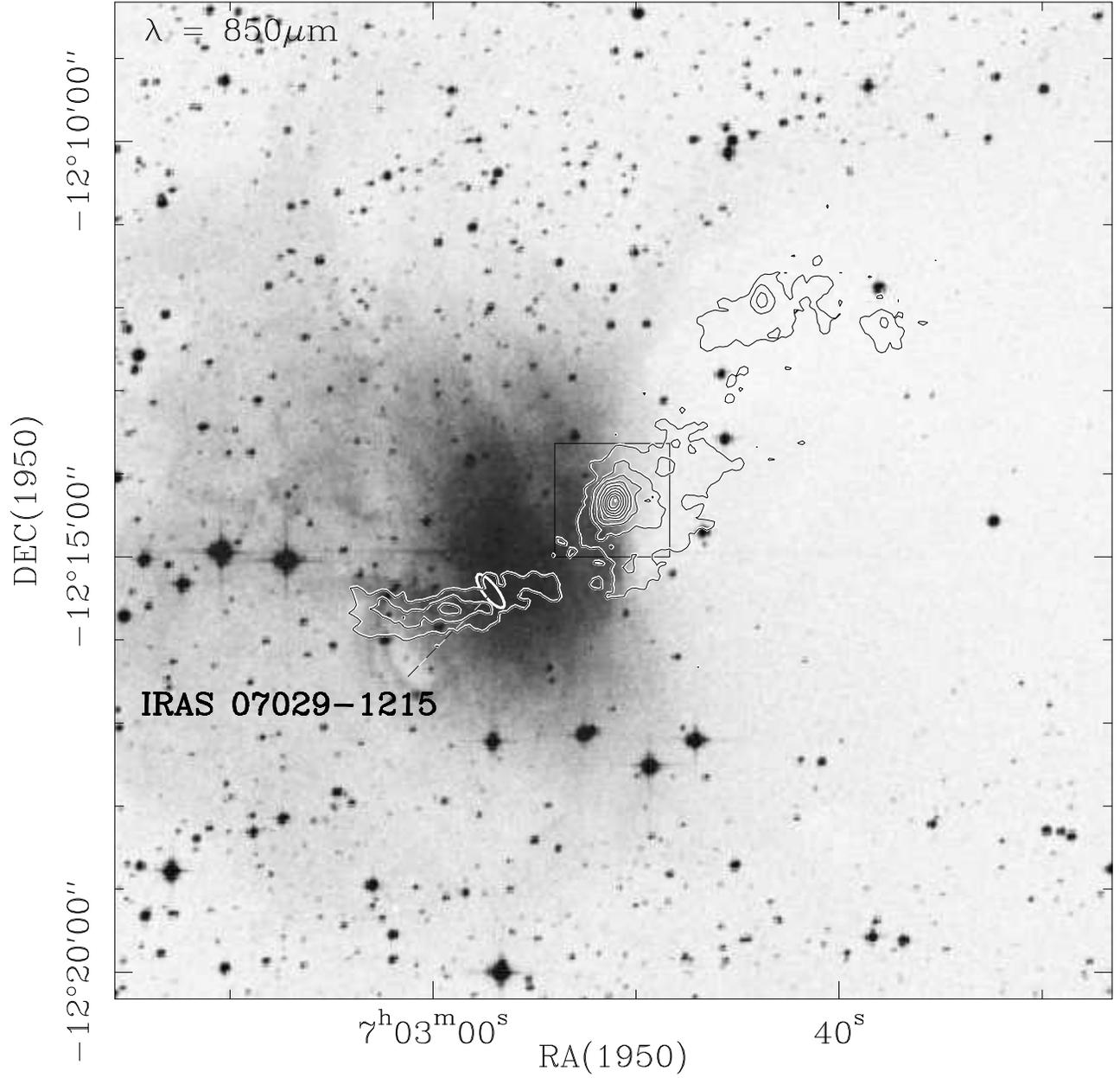}}
\caption{The surroundings of UYSO~1: POSS-I, overlaid with 850~$\mu$m continuum from the JCMT. The box indicates the FOV of Fig. 2. \label{fig1}}
\end{figure}

The POSS\footnote{The Digitized Sky Surveys were produced at the Space Telescope Science Institute under U.S.
Government grant NAGW-2166. The images of these surveys are based on photographic data obtained using the Oschin Schmidt Telescope on
Palomar Mountain and the UK Schmidt Telescope. The plates were processed into the present compressed digital form with the permission of
these institutions.} 
view in Fig.~\ref{fig1} shows the environment of IRAS~07029-1215. The overlaid SCUBA map at 850 $\mu$m displays UYSO 1 as a compact emission peak at $\alpha_{\rm (B1950.0)}=07^{\rm h}02^{\rm m}51^{\rm s} , \delta_{\rm (B1950.0)}=-12$\degree$14'26"$, between S~297
and the dark cloud. In addition to this, two dusty filaments are visible at 850~$\mu$m. The eastern filament contains the IRAS point source 
position, the western one is located inside the dark cloud.

\begin{figure}
\rotatebox{-90}{
\plotone{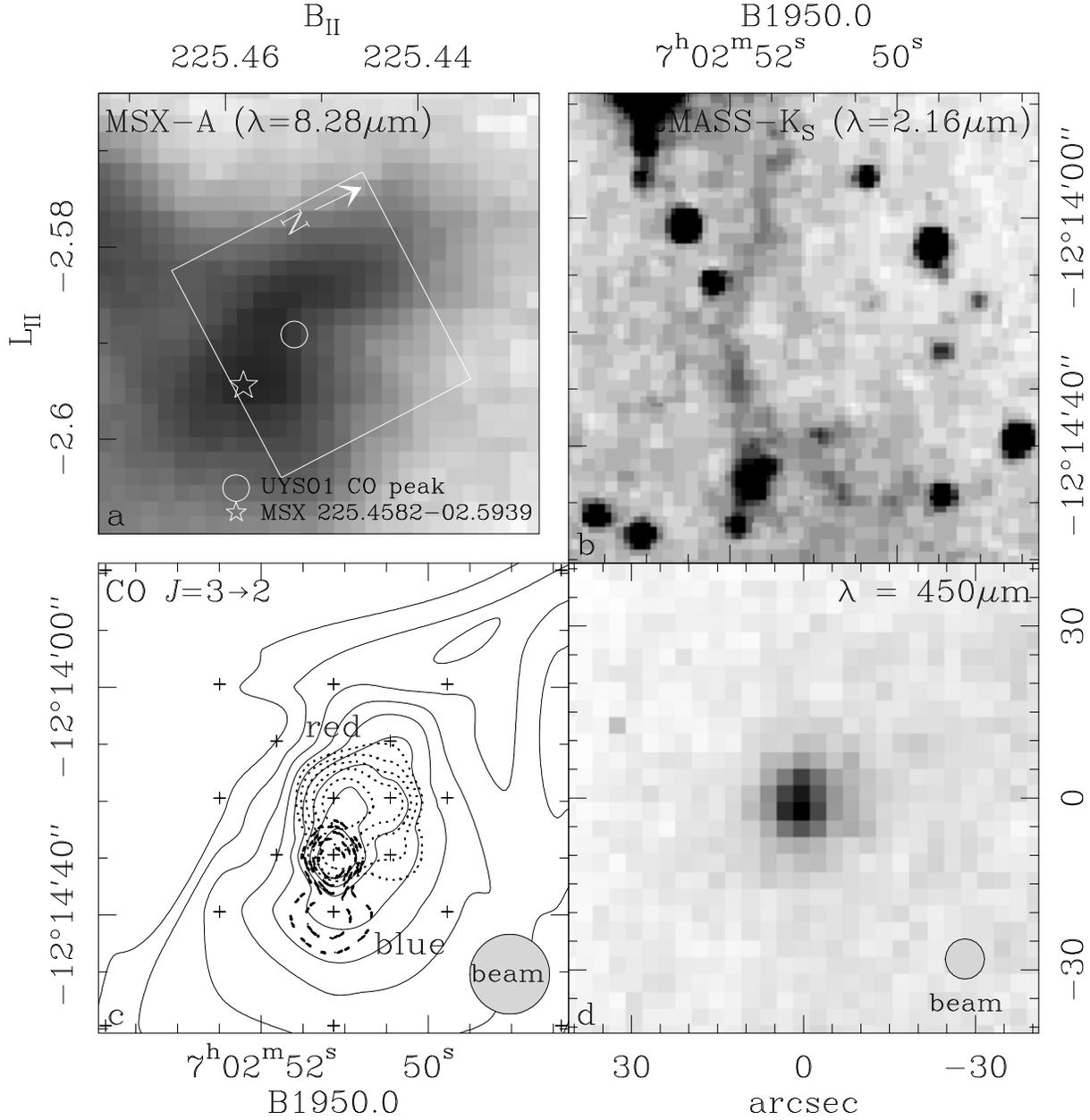}}
\caption{Close-up views of UYSO~1 in a) MSX-$A$, b) 2MASS-$K_S$, c) CO($J=3 \to 2$) and d) 450 $\mu$m continuum (both JCMT). The box in the MSX data shows the FOV of the other pictures. The integrated CO line data is shown 
in 10\% contours, the two line wings are shown down to 50\% intensity. Spectra positions are indicated as well as the beam sizes for CO and the 450~$\mu$m continuum.
UYSO~1 is located just below the center position at [0\arcsec{},-10\arcsec{}], the position of the CO peak.
\label{fig2}}
\end{figure}

UYSO 1 is still invisible at near- and 
mid-infrared wavelengths: No infrared point source in the $\lambda = $2.2..20~$\mu$m range is visible in 2MASS and MSX images (see Fig.~\ref{fig2}a,b), 
but some faint "reflected" light can be seen in the 2MASS $K_S$-band image. It cannot be ruled out, however, that this is captured
H$_2$ line emission, e.g. tracing cloud surfaces and outflows. Neither the IRAS, nor the 2MASS,
 nor the MSX infrared survey point source catalogues resulted in detections at the UYSO 1 position. There is, however, an MSX point source inside the "reflected" light south of UYSO~1.

\begin{figure}
\rotatebox{-90}{
\epsscale{0.80}
\plotone{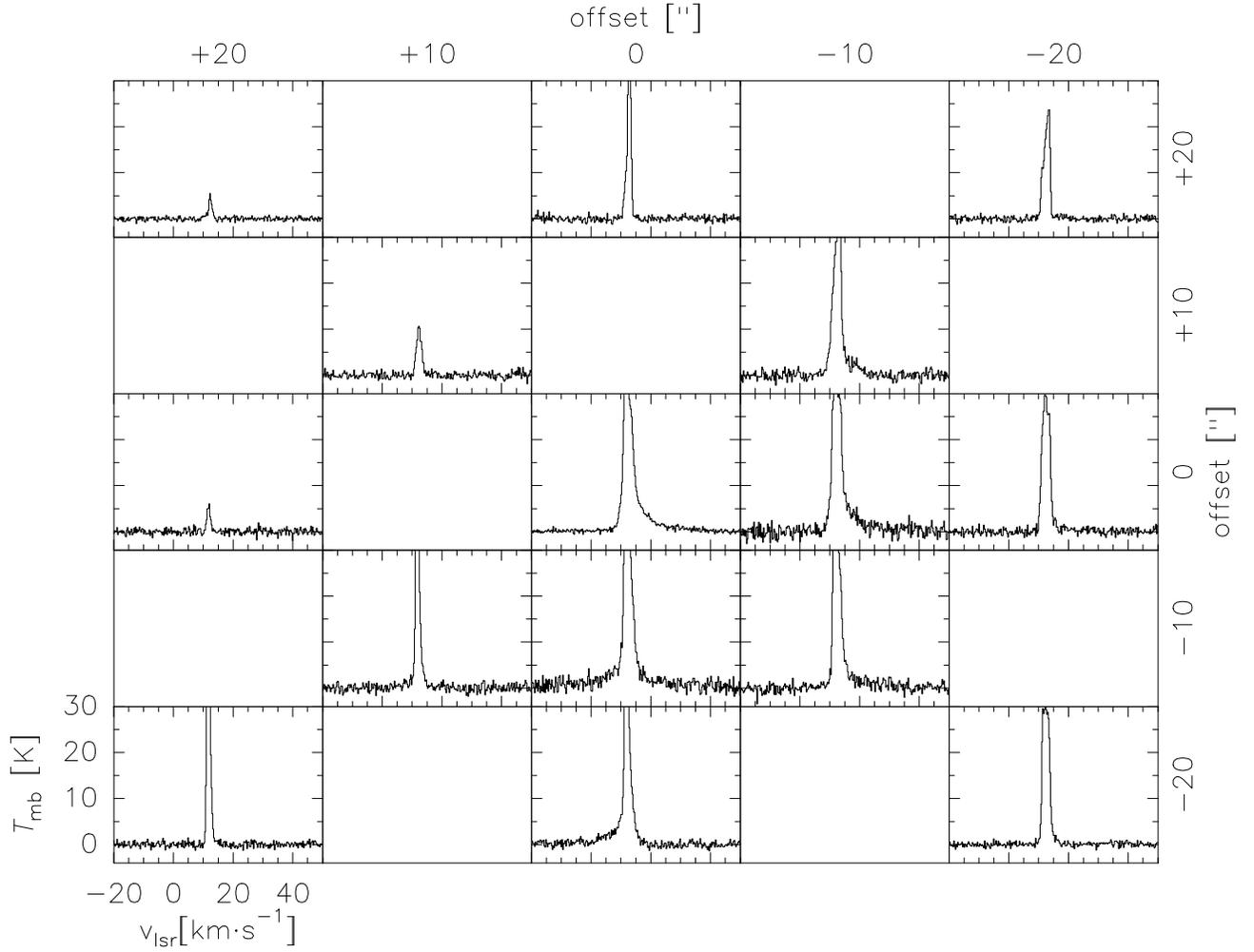}}
\caption{CO($J=3 \to 2$) spectra on the inner grid. The line maxima (all between 40~K and 50~K) were cut to allow for a better depiction of line wings.\label{fig3A}}
\end{figure}

\begin{figure}
\rotatebox{-90}{
\epsscale{0.55}
\plotone{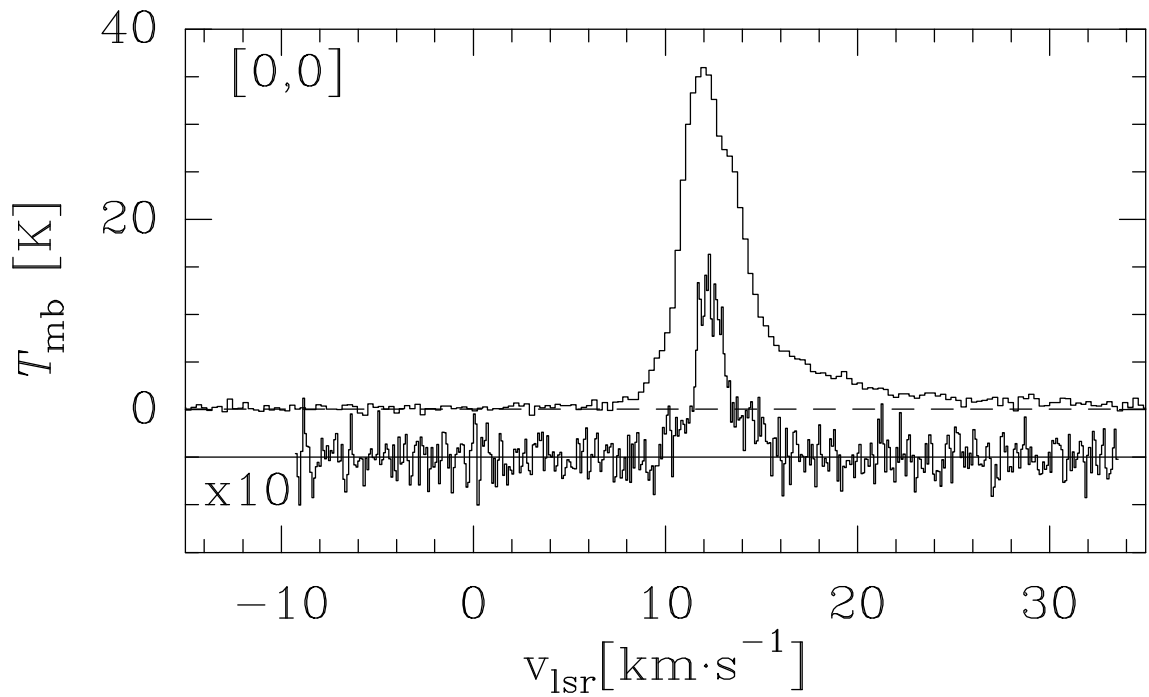}}
\caption{Comparison of the CO($J=3 \to 2$) and CS($J=5 \to 4$) lines at the respective peak positions: the CO line at [0\arcsec{},0\arcsec{}] and an addition of four CS spectra, amplified ten times for depiction, at [0\arcsec{},-6.4\arcsec{}]. Positions are in arcseconds relative to the center position of the CO grid. \label{fig3B}}
\end{figure}

The CO map in Fig.~\ref{fig2}c is a close-up view of the dense dust core. The line wings, extending from $-20$~km~s$^{-1}$ at [0",-20"] to $+40$~km~s$^{-1}$ at [0",0"] (see Fig.~\ref{fig3A}), can be clearly distinguished from well-determined baselines, extending from $-40$~km~s$^{-1}$ to $+50$~km~s$^{-1}$, respectively. Lines are narrower in off-outflow positions. Fig.~\ref{fig3B} shows a comparison of the CO($J=3 \to 2$) line profile with the CS($J=5 \to 4$) profile in the map center. CS($J=5 \to 4$) is a high-density tracer with a critical density of n$_{\rm crit}=7\times 10^6$~cm$^{-3}$.

Contrary to what could be expected from the 850~$\mu$m map, the CO data do not show a 
spherically symmetric structure. Instead, the envelope seems to be compressed towards the \HII{} region east of it. The CO peak
is roughly 10\arcsec{} off the SCUBA peak position. The line wing map displays a bipolar outflow in the NW-SE direction, its origin being UYSO~1 at raster coordinates [0\arcsec{},-10\arcsec{}], taking the emission peak of the complete integrated line as definition. The outflow is not clearly visible in the CS data, possibly due to the average density being below the CS critical densities. Lines on corresponding positions have, however, corresponding shapes.
SiO($J=2 \to 1$) could only be detected at the position
[-10\arcsec{}, 16.4\arcsec{}], in the outer regions of the redshifted outflow lobe. Possibly,
this is due to a bow shock caused by the outflow colliding with the local
interstellar medium. The detection at this position is a 6~$\sigma$ detection. With the spatial resolution of the present observations, we cannot decide whether the bipolar outflow is composed of only one or more components (cf. \citealp{beu02}).

The SCUBA 450~$\mu$m map (see Fig.~\ref{fig2}d) only shows a point source 6\arcsec{} to the north of UYSO~1, but in between the two outflow lobes.

\section{Analysis}
From the CO($J=3 \to 2$) observations, the object mass and outflow parameters can be deduced. The mass derived from an
integration of
column
densities can be compared to the virial mass. Line wing analysis delivers physical parameters of the bipolar outflow,
however still
dependent on the unknown outflow inclination angle. The distance to UYSO~1, assumed to be about the same as the distance
to IRAS~07029-1215,
was kinematically determined to be 1~kpc by \citet{wbr89}. However, while the distance to the illuminating star of S~297,
namely HD~53623, was spectroscopically determined to be 2.2~kpc $\pm$ 25\% by \citet{nec80}, Hipparcos measured a parallax for
HD~53623 of $\pi = 2.21 \pm 1.01$ mas, corresponding to a distance of $450^{+380}_{-140}$~pc \citep{esa97}. Assuming the true distance is near the
upper limit of this interval, we adopt a value of 1~kpc for the following mass estimates, in accordance with detailed studies of CMa~R1
by \citet{she99} and \citet{dez99}.

The CO data analysis is depicted in Fig.~\ref{fig2}c, in
direct comparison
to 2MASS-$K_S$\footnote{Atlas Image obtained as part of the Two Micron All Sky Survey (2MASS), a joint project of the
University of
Massachusetts and the Infrared Processing and Analysis Center/California Institute of Technology, funded by the
National Aeronautics and
Space Administration and the National Science Foundation.} data and the 450~$\mu$m continuum.

\subsection{Temperature Determination}

\begin{figure}
\rotatebox{-90}{
\epsscale{0.5}
\plotone{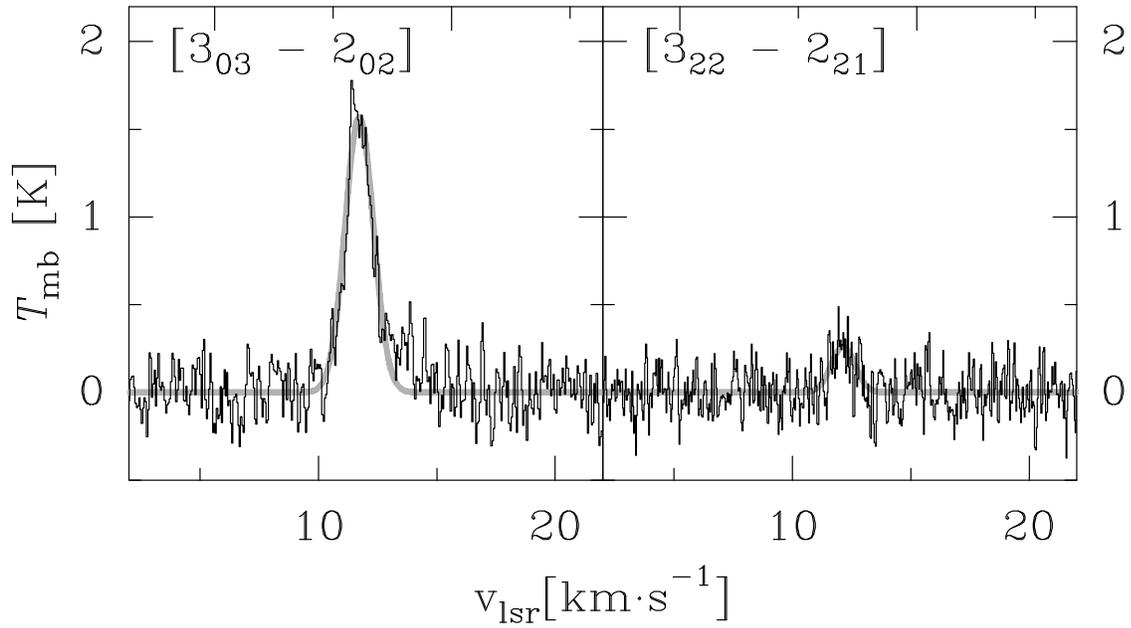}}
\caption{Temperature determination with H$_2$CO lines (see text). Gaussian fit results are shown in grey.\label{fig5}}
\end{figure}

While the main beam temperatures $T_{\rm mb}$ of the highly optically thick CO lines already confine the kinetic temperature to the 40-50~K range, another estimate was obtained from H$_2$CO($J=3_{03} \to 2_{02}$)/($J=3_{22} \to 2_{21}$) data, following \citet{man93}.
Fig.~\ref{fig5} shows the two H$_2$CO transitions for UYSO~1. The intensity ratio deduced from fitted Gaussians is 6, thus indicating a kinetic temperature of T=40$\pm 5$~K for a relatively wide range of densities.

\subsection{Mass Determination}
\label{massdet}

Assuming virial conditions, i.e. assuming that the observed line width is entirely due to motion in gravitational equilibrium, 
the total mass of 
the object can be estimated as described by \citet{hen00}. Inside a sphere of 0.14~pc radius, corresponding to the 40\% 
contour level (deconvolved with the 14\arcsec{} beam) at 
10~$\sigma$, and using the Gaussian fit result of $\Delta v=3$~km~s$^{-1}$, a virial mass of $M_{\rm vir} = 180$~M$_{\odot}$ is 
obtained.

Following \citet{hen00}, we derive hydrogen column densities from the CO($J=3\to 2$) transition,
assuming a relation of $N(\rm H_2)$~(cm$^{-2}) = 3 \times 10^{20} \int T_{\rm mb} \rm dv$. 
An underlying assumption is a conversion factor of CO($J=3\to 2$) / CO($J=1\to 0$) close to unity, as it is frequently observed in massive star-forming regions and explained by clumpy structure \citep{sto00}.
The gas mass of the cloud can then be determined as $M_{\rm gas} = 2$m$_{\rm H}\bar \mu A \langle N($H$_2)\rangle$, where $\bar \mu=1.36$ takes into account the interstellar mean
molecular weight per H atom. $A$ is the area occupied by the source.

Within the same area and to the same confidence level as the virial analysis, this leads to a different result for the cloud mass. 
With an average 
column density of $N$(H$_2$) = $3.2 \times 10^{22}$~cm$^{-2}\!\!,$ the resulting mass is $M_{\rm gas} = 40$~M$_{\odot}$.

Another estimate for the total mass has been calculated from the continuum observations at $\lambda = 850$~$\mu$m. Assuming a dust temperature of $T_{\rm d}=20$~K and optically thin conditions, $M_{\rm gas}=F_\nu \cdot d^2 / (\kappa_m(\nu) \cdot B_\nu(T_{\rm d})) \cdot [M_{\rm gas}/M_{\rm dust}]$. Using an interpolated opacity of $\kappa = 1.97$~cm$^{-2}$/g (thin ice mantles, \citealp{osh94}), the result is $M=15$~M$_{\odot}$. 
The radius of the region taken into account for this calculation is 0.11~pc (at a distance of 1~kpc), slightly smaller than the 0.14~pc used here.  Additionally, the dust continuum emission traces only the high-density central regions while the CO emission is collected along the line of sight throughout the low-density halo. Optically thick emission might further influence the estimate (see \S \ref{SEDtext}). These two points at least partly explain the discrepancy.

Thus, the mass estimates based on line and dust emission are compatible when keeping in mind their respective limitations and range between 15 and 40~M$_{\odot}$. The virial mass is much higher which is certainly caused by an overestimation of line width due to optical thickness and turbulent motion, in addition to the cloud not being in virial equilibrium.

\subsection{Outflow Properties}

The two line wings were analyzed to their respective 10\% contour levels, corresponding to a 2~$\sigma$ detection for the 
redshifted 
line wing and a 4~$\sigma$ detection for the blueshifted line wing. The masses inside the two outflow lobes are 
again calculated by 
integration of H$_2$ column densities, resulting in $M_{\rm blue} = 1.8$~M$_{\odot}$ and $M_{\rm red} = 3.6$~M$_{\odot}$, the 
total outflow mass 
thus 
being $M_{\rm out} = 5.4$~M$_{\odot}$.

The maximum projected velocities of the red and blue line wings are $v_{\rm proj,r}=27.9$~km~s$^{-1}$ and $v_{\rm proj,b}=32.1$~km~s$^{-1}$ respectively, 
the mean value being 
$v_{\rm proj}=30$~km~s$^{-1}$. Together with the known distance and the outflow radial sizes, the dynamical timescale and the 
mass loss rate can be 
obtained. With caution, the latter can be empirically converted to luminosity, and spectral type as well as mass of an assumed 
central star. All values 
discussed hereafter are summarized in Tab.~\ref{t1} for inclinations of $i=10^{\circ}\!\!,$ $i=57.3^{\circ}$ (the most 
probable value, cf. 
\citealp{bon96}) and $i=80^{\circ}\!\!.$

When the apparent outflow velocity $v_{\rm proj}$ and the apparent maximum outflow extension have been 
inclination-corrected to $v_{\rm 
out}$ and 
$R_{\rm out}$, the dynamical timescale can be calculated from $t_{\rm d}(i) = R_{\rm out}(i)/v_{\rm out}(i)$.  The 
resulting values range from $t_{\rm d}(80^{\circ}) \approx 500$~years to $t_{\rm d}(10^{\circ}) \approx 1.5 \times 10^4$~years. However, the timescale estimated this way is not a good age indicator.
Only a massive object can uphold mass outflow
rates $\dot M(i) = M_{\rm out}(i)/t_{\rm d}(i)$ ranging from $\dot M(10^{\circ}) \approx 4 \times 10^{-4}$~M$_{\odot}$/yr to $\dot M(80^{\circ}) \approx 1 \times 10^{-2}$~M$_{\odot}$/yr. 
They can be empirically transformed into the luminosity of the central object 
according to \citet{shc96}, \citet{hen00} or, based on a more
comprehensive sample, by \citet{beu02}. Tab.~\ref{t2} summarizes the resulting
spectral types, masses and luminosities (using data by \citealp{lan91}).
However, these properties of the central object can be better determined using the SED (see \S \ref{SEDtext}).

It may be interesting to consider the age of the neighbouring \HII{} region. Following \citet{cer98}, this
age can be estimated \citep{whi94}.
The size of S~297 in the optical regime with a diameter of 7' (Sharpless catalogue) corresponds to a radius 
of only about $2.1 \times 10^5$~AU, hence indicating that the nebula is rather young. Assuming a typical mean density
of $10^3$~cm$^{-3}$ and a Lyman continuum flux for a B1 star from \citet{tho84}, the resulting age is approximately
$7 \times 10^5$~yrs. 

Regardless of the uncertainties in these age estimates, the \HII{} region is likely to be older than UYSO~1. Possibly, this is an example of induced star formation, although this would need kinematic confirmation.

\subsection{Spectral Energy Distribution}
\label{SEDtext}
While the two measured continuum fluxes at 450~$\mu$m and 850~$\mu$m determine the cold dust emission, more 
data had to be 
complemented before the spectral energy distribution could be tentatively compared to modeling results. Given the aim of 
determining the luminosity 
and mass of the central object as well as having a comparison to the mass determination, upper flux limits for short 
wavelengths helped 
pinning down physical properties of the central object. As main support besides the JCMT data, IRAS fluxes were used, determined at the position of UYSO~1 within beam sizes
from the IRAS 
atlas maps. UYSO~1 is unresolved also on IRAS HIRES maps. Additionally, data for the definitely brighter nearby MSX point source G225.4582-02.5939 and 2MASS detection limits set further 
constraints.

Integrating the SED and using the flux limits at infrared wavelengths as the actual flux values leads to a luminosity of $<1900$~L$_{\odot}$, assuming spherical symmetry.
A modified blackbody accounting for the dust optical depth was fitted to the data: $S_\nu = B_\nu(T) (1-\exp(-\tau_\nu)) \Delta\Omega$, with the beam size $\Delta\Omega$ and with dust opacities $\kappa_\nu \propto \nu^\beta$ (see e.g. \citealp{hil83}). A good fit is obtained for $T=45$~K and $\beta=2$, however fits with temperatures between 40 and 50~K are also fitting the data within the uncertainties.
With the aim of checking the consistency of existing data, the continuum 
emission was then modeled using the radiative transfer code developed by 
\citet{man98}. This is an accelerated version of the 2-D ray-tracing code 
developed for radiative transfer in disk configurations by \citet{men97}. Due 
to insufficient parameter constraints, the code was used in its 1D version only.

The limiting outer radius of the model could be estimated from the CO data to
be $3.2 \times 10^4$~AU, taking the same area as for the mass estimation. Assuming an exponential of the density 
distribution of 2.0 with a dust-to-gas ratio of 1:150, the model 
uses silicate optical data from \citet{dor95}, while the properties of the
carbonaceous dust are from \citet{jae98} [1000~K data]. The ratio of silicates to carbon is
3:2. The inner model radius is determined by the dust sublimation temperature
of 1500~K. 

\begin{figure}
\rotatebox{-90}{
\epsscale{0.7}
\plotone{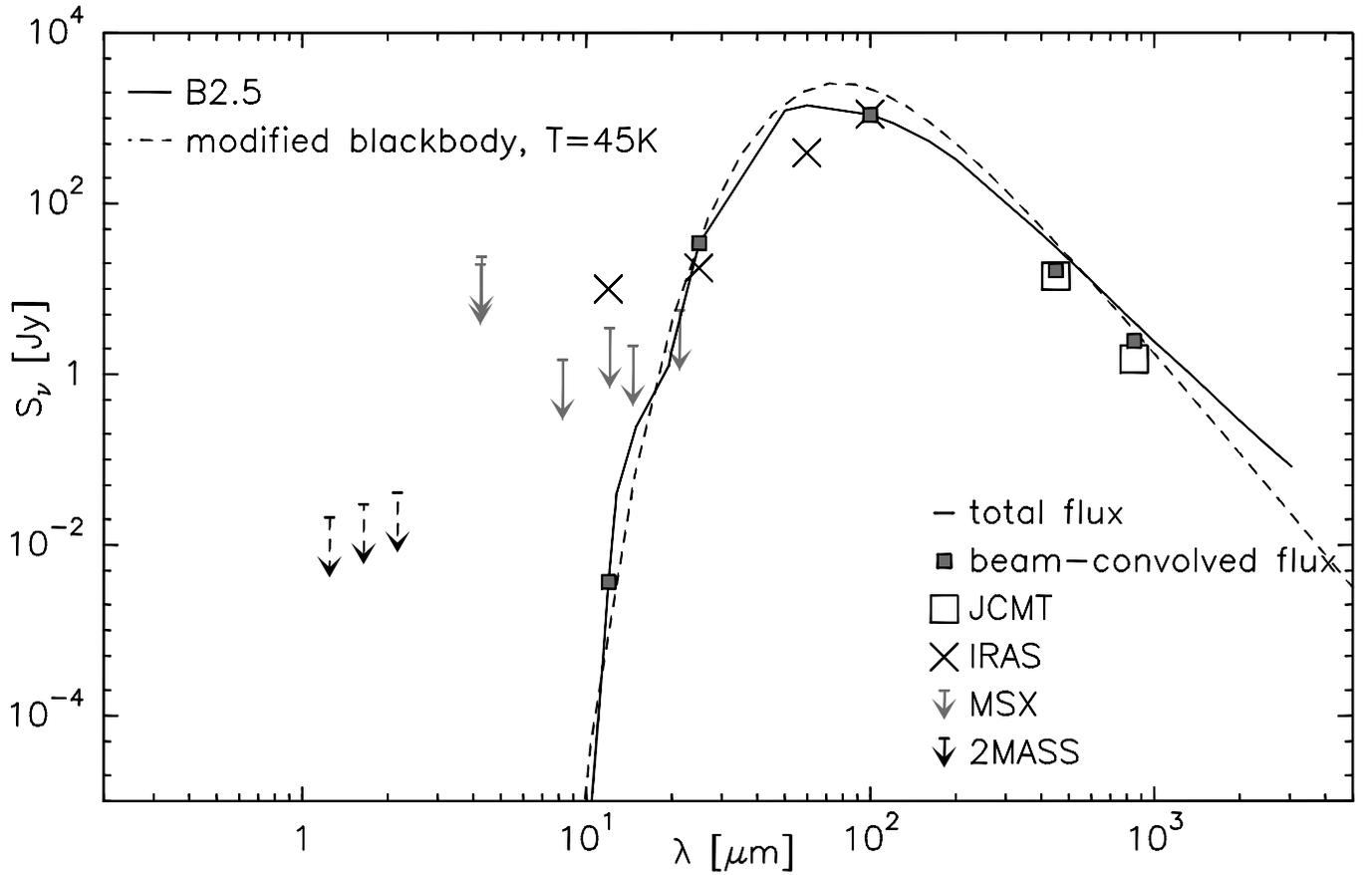}}
\caption{Modeling results from the Manske code. The straight line shows the total flux (displayed
for $\tau_\lambda \lesssim 1000$), boxes
indicate beam-convolved fluxes. The dashed line corresponds to a modified
blackbody at T=45~K (with $\beta=2$)\label{fig4}}
\end{figure}

Fig.~\ref{fig4} shows the results for an envelope mass of 
$M_{\rm env} = 30$~M$_{\odot}$ and an assumed single 
central object in the B2.5 class. The optical depth at $\lambda = 500$~nm for the simulation runs is $\tau_{500 \rm nm} \approx 6400$. Since the code becomes inaccurate for high optical depths, the computed SED is arbitrarily shown for $\tau \lesssim 1000$, corresponding to $\lambda > 10$~$\mu$m. At $\lambda = 850$~$\mu$m, the optical depth is $\tau_{850 \rm \mu m} \approx 1$. Thus, the mass estimate from the dust emission in \S \ref{massdet} is rather a lower limit because of contribution from optically thick emission.
  
Both the simulation and the blackbody fit indicate the equivalent of an early B star as the central object. This finding is compatible with the result of the empirical outflow analysis for $i=57.3^{\circ}$.
The envelope mass in the simulation is compatible with the mass estimate from the CO($J=3 \to 2$) data (40~M$_{\odot}$), 
especially when keeping in mind that the
modeled gas density only reaches $10^4$~cm$^{-3}$ at its outer radius, thus
allowing for a slightly higher envelope mass when integrating further outwards. 

\section{Conclusions}

The emerging picture of UYSO~1 drawn from both mass estimates and subsequent empirical relations 
as well as SED considerations thus is that of an early B star surrounded by an 
envelope of about 30-40~M$_{\odot}$. The object seems to be in a particularly early evolutionary state, but already drives a bipolar outflow with a total mass of M$_{\rm outflow} = 5.4$~M$_{\odot}$. To our knowledge, there are only very few comparable sources that are undetected at infrared wavelengths (see e.g. \citealp{hun98}, \citealp{bdm99} and \citealp{fon03}).
With additional high-resolution observations, we hope to be able to specify these conclusions and learn more about this evolutionary state. Possibly, parameters of a putative disk around UYSO~1 could be derived then.

\acknowledgments

The authors are grateful to Remo Tilanus for his support during the JCMT observing run, as well as to R. Mauersberger for carrying out the DDT observations at the IRAM 30m telescope in service mode.

This research has made use of the SIMBAD database, operated at CDS, Strasbourg, France.

\clearpage

\begin{table}
\begin{center}
\caption{Heterodyne observations of UYSO~1\label{t0}}
\begin{tabular}{lrrrcc}
\tableline\tableline
Line& Frequency& Telescope& Date& beam size& $\eta_{\rm mb}$\\
\tableline
CO($J=3 \to 2$)& 345.79 GHz& JCMT& Oct 99& 14"& 0.63\\
CS($J=2 \to 1$)&  97.98 GHz& IRAM 30m& Mar 03& 25"& 0.78\\
CS($J=5 \to 4$)& 244.94 GHz& IRAM 30m& Mar 03& 10"& 0.50\\
SiO($J=2 \to 1$)& 86.84 GHz& IRAM 30m& Mar 03& 29"& 0.78\\
H$_2$CO($J=3-2$)\tablenotemark{a}&218.34 GHz& IRAM 30m& Mar 03& 12"& 0.56\\
\tableline
\end{tabular}
\tablenotetext{a}{H$_2$CO($J=3_{03} \to 2_{02}$) and ($J=3_{22} \to 2_{21}$)}
\end{center}
\end{table}

\begin{table}
\begin{center}
\caption{UYSO~1 outflow properties\label{t1}}
\begin{tabular}{ccccc}
\tableline\tableline
i& v$_{\rm out}$ & R$_{\rm out}$ & t$_{\rm d}$ & $\dot {\rm{M}}$\\
\tableline
$10^{\circ}$ & 30.5 km s$^{-1}$ & $1.4 \times 10^{13}$ km & 14600 yrs & $ 3.7 \times 10^{-4} $ M$_{\odot}$/yr\\
$57^{\circ}\!\!.3$ & 55.5 km s$^{-1}$ & $2.9 \times 10^{12}$ km & 1700 yrs & $3.3 \times 10^{-3} $ M$_{\odot}$/yr\\
$80^{\circ}$ & 172.7 km s$^{-1}$ & $2.4 \times 10^{12}$ km & 450 yrs & $1.2 \times 10^{-2} $ M$_{\odot}$/yr\\
\tableline
\end{tabular}
\end{center}
\end{table}


\begin{table}
\begin{center}
\caption{Outflow-inferred properties of assumed central star\label{t2}}
\begin{tabular}{cccc}
\tableline\tableline
i & L & spec. type & M$_{\rm star}$\\
\tableline
$10^{\circ}$ & $10^3 $ L$_{\odot}$ & B4 & 6.5 M$_{\odot}$\\
$57^{\circ}\!\!.3$ & $10^4 $ L$_{\odot}$ & B1 & 13 M$_{\odot}$\\
$80^{\circ}$ & $10^5 $ L$_{\odot}$ & O9 & 19 M$_{\odot}$\\
\tableline
\end{tabular}
\end{center}
\end{table}

\clearpage


\begin{thebibliography}{}
\bibitem[Bernard, Dobashi, \& Momose(1999)]{bdm99} Bernard, J. P., \& Dobashi, K., Momose, M. 1999, A\&A, 350, 197
\bibitem[Beuther et al.(2002)]{beu02b} Beuther, H., Schilke, P., Menten, K. M., Motte, F., Sridharan, T. K., \& Wyrowski, F. 2002, ApJ, 566, 945
\bibitem[Beuther et al.(2002b)]{beu02} Beuther, H., Schilke, P., Sridharan, T.K., Menten, K.M., Walmsley, C.M., \& Wyrowski, F. 2002, A\&A, 383, 892
\bibitem[Bonnell, Bate, \& Zinnecker(1998)]{bon98} Bonnell, I.A., Bate, M.R., \& Zinnecker, H.  1998, MNRAS, 298, 93
\bibitem[Bontemps, Ward-Thompson, \& Andr\'e(1996)]{bon96} Bontemps, S., Ward-Thompson, D., \& Andr\'e, P. 1996, A\&A, 314, 477
\bibitem[Cernicharo et al.(1998)]{cer98} Cernicharo, J., Lefloch, B., Cox, P., Cesarsky, D., Esteban, C., Yusef-Zadeh, F., M\'endez, D.I., Acosta-Pulido, J., Garcia L\'opez, R.J., \& Heras, A. 1998, Science, 282, 462
\bibitem[de Zeeuw, Hoogerwerf, \& de Bruijne(1999)]{dez99} de Zeeuw, P.T., Hoogerwerf, R., \& de Bruijne, J.H.J. 1999, AJ 117, 354
\bibitem[Dorschner et al.(1995)]{dor95} Dorschner, J., Begemann, B., Henning, Th., J\"ager, C., \& Mutschke, H. 1995, A\&A, 300, 503
\bibitem[Emerson (1995)]{eme95} Emerson, D. T. 1995, in ASP Conf. Ser., Vol. 75, Multi-feed Systems for Radio Telescopes, ed. D. T. Emerson, J. M. Payne, (San Francisco: ASP)
\bibitem[ESA(1997)]{esa97} ESA 1997, The Hipparcos and Tycho Star Catalogues, ESA SP-1200
\bibitem[Evans et al.(2002)]{eva02} Evans II, N.J., Shirley, Y.L., Mueller, K.E., \& Knez, C. 2002, in ASP Conf.Ser.267, Hot Star Workshop III: The Earliest Phases of Massive Star Birth, ed. P.A. Crowther (San Francisco: ASP)
\bibitem[Fontani et al.(2003)]{fon03} Fontani, F., Cesaroni, R., Testi, L., Walmsley, C.M., Molinari, E., Neri, R., Shepherd, D., Brand, J., Palla, F., \& Zhang, Q., A\&A, \textsl{in press}
\bibitem[Henning et al.(1992)]{hen92} Henning, Th., Cesaroni, R., Walmsley, M., \& Pfau, W. 1992, \aaps, 93, 525
\bibitem[Henning et al.(2000)]{hen00} Henning, Th., Schreyer, K., Launhardt, R., \& Burkert, A. 2000, A\&A, 353, 211
\bibitem[Hildebrand(1983)]{hil83} Hildebrand, R.H. 1983, \qjras, 24, 267
\bibitem[Houk \& Smith-Moore(1988)]{hou88} Houk, N., \& Smith-Moore, M. 1988, Michigan catalogue of Two-Dimensional spectral types for the HD stars,vol. 4 : -26 to -12 degrees, Michigan Spectral Survey (Ann Arbor, Dept. of Astronomy, Univ. Michigan)
\bibitem[Hunter et al.(1998)]{hun98} Hunter, T.R., Neugebauer, G., Benford, D.J., Matthews, K., Lis, D.C., Serabyn, E., \& Phillips, T.G. 1998, ApJ, 493, L97
\bibitem[Hunter et al.(2000)]{hun00} Hunter, T. R., Churchwell, E., Watson, C., Cox, P., Benford, D. J., \& Roelfsema, P. R. 2000, AJ, 119, 2711
\bibitem[J\"ager et al.(1998)]{jae98} J\"ager, C., Mutschke, H., Dorschner, J., \& Henning, Th. 1998, A\&A, 332, 291
\bibitem[Jijina \& Adams(1996)]{jia96} Jijina, J., \& Adams, F.C. 1996, ApJ, 462, 874
\bibitem[Lang(1991)]{lan91} Lang, K.R. 1991, Astrophysical Data: Planets and Stars (Berlin: Springer)
\bibitem[Mangum \& Wootten(1993)]{man93} Mangum, J.G., \& Wootten, A. 1993, ApJS, 89, 123
\bibitem[Manske, Henning, \& Men'shchikov(1998)]{man98} Manske, V., Henning, Th., \& Men'shchikov, A.B. 1998, A\&A, 331, 52
\bibitem[Men'shchikov \& Henning(1997)]{men97} Men'shchikov, A. B., \& Henning, Th. 1997, A\&A, 318, 879
\bibitem[Mueller et al.(2002)]{mue02} Mueller, K.E., Shirley, Y.L., Evans, N.J., II, \& Jacobson, H.R. 2002, ApJS, 143, 469
\bibitem[Neckel, Klare, \& Sarcander(1980)]{nec80} Neckel, Th., Klare, G., \& Sarcander, M. 1980, A\&ASS, 42, 251
\bibitem[Ossenkopf \& Henning(1994)]{osh94} Ossenkopf, V., \& Henning, Th., A\&A, 291, 3
\bibitem[Posselt(2003)]{pos03} Posselt, B. 2003, diploma thesis, Friedrich Schiller University, Jena, Germany
\bibitem[Rohlfs \& Wilson(2000)]{rwi00} Rohlfs, K., \& Wilson, T.L., 2000, Tools of Radio Astronomy (Berlin: Springer)
\bibitem[Shepherd \& Churchwell(1996)] {shc96} Shepherd, D.S., \& Churchwell, E. 1996, ApJ, 472, 225
\bibitem[Shevchenko et al.(1999)]{she99} Shevchenko, V.S., Ezhkova, O.V., Ibrahimov, M.A., van den Ancker, M.E., \& Tjin A Djie, H.R.E. 1999, MNRAS, 310, 210
\bibitem[St\"orzer et al.(2000)]{sto00} St\"orzer, H., Zielinsky, M., Stutzki, J., \& Sternberg, A. 2000, A\&A, 358, 682 
\bibitem[Thompson(1984)]{tho84} Thompson, R.I. 1984, ApJ, 283, 165
\bibitem[Whitworth et al.(1994)]{whi94} Whitworth, A.P., Bhattal, A.S., Chapman, S.J., Disney, M.J., \& Turner, J.A. 1994, MNRAS, 268, 291
\bibitem[Wouterloot \& Brand(1989)]{wbr89} Wouterloot, J.G.A., \& Brand, J. 1989, A\&AS, 80, 149

\end{thebibliography}
\end{document}